%% file: main.tex
\documentclass[conference]{IEEEtran}
\IEEEoverridecommandlockouts
\usepackage{cite}
\usepackage{amsmath,amssymb,amsfonts}
\usepackage{algorithmic}
\usepackage{graphicx}
\usepackage{textcomp}
\usepackage{xcolor}
\def\BibTeX{{\rm B\kern-.05em{\sc i\kern-.025em b}\kern-.08em
    T\kern-.1667em\lower.7ex\hbox{E}\kern-.125emX}}
\usepackage{comment}  
\usepackage{soul}
\usepackage{booktabs}
\usepackage{multirow}
\usepackage{numprint}
\usepackage{todonotes}
\usepackage{url}
\usepackage{hyperref}
\usepackage{relsize}
\usepackage{cancel}
\usepackage{amsmath}
\usepackage{tikz}
\newcommand*\circled[1]{\tikz[baseline=(char.base)]{
            \node[shape=circle,draw,inner sep=1pt,font=\sffamily\footnotesize] (char) {\textbf{#1}};}}

\usepackage{listings}
\lstdefinelanguage
   [x64]{Assembler}     
   [x86masm]{Assembler} 
   {morekeywords={CDQE,CQO,CMPSQ,CMPXCHG16B,JRCXZ,LODSQ,MOVSXD, %
                  POPFQ,PUSHFQ,SCASQ,STOSQ,IRETQ,RDTSCP,SWAPGS, %
                  rax,rdx,rcx,rbx,rsi,rdi,rsp,rbp, 
                  r8,r8d,r8w,r8b,r9,r9d,r9w,r9b, %
                  r10,r10d,r10w,r10b,r11,r11d,r11w,r11b, %
                  r12,r12d,r12w,r12b,r13,r13d,r13w,r13b, %
                  r14,r14d,r14w,r14b,r15,r15d,r15w,r15b,}} 

\usepackage{xcolor}

\definecolor{codegreen}{rgb}{0,0.6,0}
\definecolor{codegray}{rgb}{0.5,0.5,0.5}
\definecolor{codepurple}{rgb}{0.58,0,0.82}
\definecolor{backcolour}{rgb}{0.95,0.95,0.92}

\lstdefinestyle{mystyle}{
    backgroundcolor=\color{backcolour},   
    commentstyle=\color{codegreen},
    keywordstyle=\color{blue},
    numberstyle=\tiny\color{codegray},
    stringstyle=\color{codepurple},
    basicstyle=\ttfamily\footnotesize,
    breakatwhitespace=false,         
    breaklines=true,                 
    captionpos=b,                    
    keepspaces=true,                 
    numbers=left,                    
    numbersep=5pt,                  
    showspaces=false,                
    showstringspaces=false,
    showtabs=false,                  
    tabsize=2,
    columns=fullflexible,
}
\newcommand{\approach}[1]{\emph{EVIL}}
\usepackage[all=normal,floats,paragraphs,mathspacing,wordspacing,charwidths,indent,lists]{savetrees}

\begin{document}

\title{\textit{EVIL}: Exploiting Software via Natural Language}

\author{
    \IEEEauthorblockN{Pietro Liguori\IEEEauthorrefmark{1}, Erfan Al-Hossami\IEEEauthorrefmark{2}, Vittorio Orbinato\IEEEauthorrefmark{1},\\ Roberto Natella\IEEEauthorrefmark{1}, Samira Shaikh\IEEEauthorrefmark{2}, Domenico Cotroneo\IEEEauthorrefmark{1}, and Bojan Cukic\IEEEauthorrefmark{2}}
    \IEEEauthorblockA{\IEEEauthorrefmark{1}University of Naples Federico II, Naples, Italy
    \\\{pietro.liguori, vittorio.orbinato, roberto.natella, cotroneo\}@unina.it}
    \IEEEauthorblockA{\IEEEauthorrefmark{2}University of North Carolina at Charlotte, Charlotte, NC
    \\\{ealhossa, samirashaikh, bcukic\}@uncc.edu}
}

\maketitle
\thispagestyle{plain}
\pagestyle{plain}

\begin{abstract}
Writing exploits for security assessment is a challenging task. The writer needs to master programming and obfuscation techniques to develop a successful exploit. 
To make the task easier, we propose an approach (\approach{}) to automatically generate exploits in assembly/Python language from descriptions in natural language. 
The approach leverages Neural Machine Translation (NMT) techniques and a dataset that we developed for this work. 
We present an extensive experimental study to evaluate the feasibility of \approach{}, using both automatic and manual analysis, and both at generating individual statements and entire exploits. 
The generated code achieved high accuracy in terms of syntactic and semantic correctness.
\end{abstract}

\begin{IEEEkeywords}
Automatic Exploit Generation, Neural Machine Translation, Software Exploits, Shellcode, Encoder, Decoder
\end{IEEEkeywords}

\section{Introduction}
\label{sec:introduction}
\input{tex/introduction}

\section{Background}
\label{sec:background}
\input{tex/background}

\section{Proposed Methodology}
\label{sec:methodology}
\input{tex/methodology}

\section{Model Training}
\label{sec:dataset}
\input{tex/dataset}

\section{Experimental Evaluation}
\label{sec:experiments}
\input{tex/experiments}

\section{Qualitative Analysis}
\label{sec:qualitative}
\input{tex/qualitative}

\section{Related Work}
\label{sec:related}
\input{tex/related}

\section{Ethical Considerations}
\label{sec:ethics}
\input{tex/ethics}

\section{Conclusion}
\label{sec:conclusion}
\input{tex/conclusion}

\section*{Acknowledgment}
This work has been partially supported by the University of Naples Federico II in the frame of the Programme F.R.A., project id OSTAGE, the Italian Ministry of University and Research (MUR) under the programme ``PON Ricerca e Innovazione 2014-2020 – Dottorati innovativi con caratterizzazione industriale'' and Cisco Systems Inc., project id BOTITS. 

\IEEEtriggeratref{62}
\bibliographystyle{IEEEtran}
\bibliography{bibliography}

\end{document}


\title{\textit{EVIL}: Exploiting Software via Natural Language}

\maketitle
\thispagestyle{plain}
\pagestyle{plain}

\appendix
\label{sec:appendix}
\input{tex/appendix_table}

%% file: tex/introduction.tex
In the context of software security, a solid understanding of offensive techniques is increasingly important \cite{avgerinos2011aeg,avgerinos2014automatic}. Well-intentioned actors, such as penetration testers, ethical hackers, researchers, and computer security teams are engaged in developing exploits, referred to as \textit{proof-of-concept} (POC), to reveal security weaknesses within the software. Offensive security helps us understand how attackers take advantage of vulnerabilities and motivates vendors and users to patch them to prevent attacks \cite{arce2004shellcode}. 
Among software exploits, code-injection attacks are the trickiest. They allow the attacker to inject and execute arbitrary code on the victim system. Since the injected code frequently launches a command shell, the hacking community refers to the payload portion of a code-injection attack as a \textit{shellcode} \cite{mason2009english}. 

Writing code injection exploits is a challenging task since it requires significant technical skills. Shellcodes are typically written in assembly language, affording the attacker full control of the memory layout and CPU registers to attack low-level mechanisms (e.g., heap metadata and stack return addresses) not otherwise accessible through high-level programming languages. 
Another challenge for shellcodes is modern antivirus (AV) and intrusion detection systems (IDS), which actively look for malicious payloads to block attacks. 
To elude detection, shellcode writers weaponize their shellcode by implementing an encoding/decoding strategy. In other words, writers have to develop \emph{encoders} (typically, using Python) to obfuscate the original shellcode without altering its functionality, and \emph{decoders} (typically in assembly language, as the shellcode) to revert to the payload once it is loaded (and then executed) on the victim system.



In this work, we propose an approach, \approach{} (Exploiting software VIa natural Language), for exploit writing based on natural language processing. The approach aims to
support both beginners and experienced researchers, by making exploits easier to create and flattening the learning curve. 
In \approach{}, a machine learning system learns about exploit writing from a dataset, containing both real exploits and their description in the English language. Then, the writer describes the exploit using the English language and lets the machine learning system translate the description into assembly and Python code. \approach{} leverages recent advances in \emph{neural machine translation} (NMT) to automatically generate code from natural language descriptions using recurrent neural networks. 
NMT has emerged as a promising machine translation approach, and it is widely recognized as the state-of-the-art method for the translation of different languages \cite{wu2016google,bojar2016findings}. 
NMT has been adopted in many different areas, to generate programs in the Python language \cite{yin2017syntactic, DBLP:journals/corr/LingGHKSWB16}, OS commands for the UNIX Bash shell \cite{lin2017program,lin2018nl2bash}, commit messages for version control \cite{jiang2017automatically,liu2018neural}, code completion \cite{ciniselli2021empirical}, test cases from security requirements \cite{mai2018natural}, and more. However, NMT techniques have not heretofore been applied in the field of software security in the manner described in our approach.

Our work provides three key contributions:

\begin{itemize}
    \item We release a substantive dataset\footnote{The dataset and the code to reproduce the experiments are publicly available here: \url{https://github.com/dessertlab/EVIL}} containing exploits collected from shellcode databases and their descriptions in the English language. The dataset includes both assembly code (i.e, shellcodes and decoders) and Python code (i.e., encoders). Such data is valuable to support research in machine translation for security-oriented applications since the techniques are data-driven. 
    \item We propose a new approach that applies NMT techniques to automatically generate exploits, including both Python code for encoding the payload, and assembly code for decoding the payload and for the actual shellcode, based on their description in the English language. 
    \item We perform an extensive experimental study, to evaluate the feasibility of the proposed approach at generating real exploits. To this aim, we propose new metrics that go beyond evaluating the translation of single lines of code \cite{DBLP:journals/corr/LingGHKSWB16,yin2018tranx,yin2019reranking,Xu2020IncorporatingEK}, but encompass entire exploits as a whole. The generated code achieved high accuracy in terms of syntactic and semantic correctness.
\end{itemize}

In the following, Section~\ref{sec:background} introduces background concepts; Section~\ref{sec:methodology} presents the proposed approach; Section~\ref{sec:dataset} describes the dataset; Section~\ref{sec:experiments} experimentally evaluate the approach; Section~\ref{sec:qualitative} describes the approach further through selected examples; Section~\ref{sec:related} discusses related work; Section~\ref{sec:ethics} discusses the ethical considerations; Section~\ref{sec:conclusion} concludes the paper. 

%% file: tex/background.tex
Statistics from the Common Vulnerabilities and Exposures (CVE) database show that code-injection vulnerabilities increased dramatically during recent years \cite{management2013software,management2016identity}.  
Code-injection attacks deliver and run arbitrary code on victims’ machines, enabling unauthorized access and control of system resources, applications, and data \cite{snow2011shellos}.
Moreover, code-injection attacks have become more and more sophisticated, including techniques such as return-oriented programming, heap spraying, and format string attacks.

A shellcode is a list of machine code instructions, to be loaded in a vulnerable application at runtime. 
The classic way to develop shellcodes is to write them using the assembly language, and by using an assembler to turn them into \emph{opcodes} (operation codes, i.e., a machine language instruction in binary format, to be decoded and executed by the CPU) \cite{foster2005sockets,megahed2018penetration}. 
Listing~\ref{list:shellcode} shows an example of shellcode\footnote{Shellcode collected from \url{https://www.exploit-db.com/shellcodes/47890}} in assembly for the 32-bit Intel Architecture, which runs the \texttt{/bin/sh} command to spawn a shell on the Linux OS. Note that the shellcode strictly depends on the OS and the CPU architecture, thus it must be tailored for the target system. 
Listing~\ref{list:shellcode} also shows the shellcode as binary opcodes (lines 17-18), where each \textit{\textbackslash x} followed by two hex digits represents one byte of the payload \cite{koziol2004shellcoder}. Shellcodes typically range between few bytes to hundreds of bytes. 
Other objectives of shellcodes include killing or restart other processes, causing a denial-of-service (e.g., a fork bomb), leaking secret data, etc.

\begin{figure}[t]
\begin{minipage}{\linewidth}
\begin{lstlisting}[caption={Example of shellcode to spawn a shell on Linux x32 systems.},label={list:shellcode}]
global _start
section .text
_start:
        xor     eax, eax
        push    eax
        push    0x68732f2f
        push    0x6e69622f
        mov     ebx, esp
        push    eax
        mov     edx, esp
        push    ebx
        mov     ecx, esp
        mov     al, 11
        int     0x80
        
------------------ Binary opcodes ------------------
"\x31\xc0\x50\x68\x2f\x2f\x73\x68\x68\x2f\x62\x69\
x6e\x89\xe3\x50\x89\xe2\x53\x89\xe1\xb0\x0b\xcd\x80"
\end{lstlisting}
\end{minipage}
\end{figure}



The plain shellcodes contain explicit information of the malicious action the attacker aims to take. For example, the shellcode in Listing~\ref{list:shellcode} contains the values \textit{68 73 2f 2f} and \textit{6e 69 62 2f}, which are the hexadecimal representation of the strings \textit{//sh} and \textit{/bin} in reverse order (since the target CPU is little-endian).  Therefore, it is easy for AV and IDS software to block the execution of this shellcode. 
To overcome security protections, exploit writers adopt encoding techniques, which convert the original shellcode into a new, functionally equivalent one, but more difficult to block \cite{erickson2008hacking,regalado2015gray}. 
Encoders are programs written in a high-level language, most often in Python (e.g., over the $80\%$ of the encoders on the popular Exploit Database \cite{exploitdb} are developed in Python language), and apply mathematical operations (as in symmetric key cryptography) on the binary opcodes to generate new ones, and which append additional opcodes for the decoder. Afterward, when this attack payload is injected and executed by the victim system, it decodes itself to obtain the original shellcode. Since the decoder is part of the attack payload, it is developed using the assembly language.

Encoders and decoders are not just used to obfuscate the original payload from AV and IDS. Encoding schemes are used also to eliminate ``bad bytes'' from the payload. For example, since a null byte is considered as a terminator character for strings, all the bytes of the payload following the null bytes are not processed. Accordingly, the shellcodes must be null-free or zero-free, i.e., they can not contain any null bytes. 
Moreover, most vulnerabilities impose restrictions on the quantity of data that can be injected. Therefore, another use of encoders is to optimize the shellcode to decrease its size.

As security attacks and defenses evolve with technology, security researchers have been developing new encoding techniques. Among recent studies, Geczi et al. \cite{geczi2018automatic} described a technique that converts x86 assembly code, such that the resulting object code only contains printable characters. Similarly, Patel et al. \cite{patel2020automatic} developed a new encoding scheme to produce printable shellcodes but in a more compact and reduced size. 
The penetration testing tool Metasploit \cite{metasploit} also provides a method, named \textit{sub encoder}, to convert any sequence of binary data into ASCII characters that, when interpreted by an Intel CPU, will decode the original sequence and execute it.

%% file: tex/methodology.tex
\begin{figure*}[ht]
\centering
\includegraphics[width=0.8\linewidth]{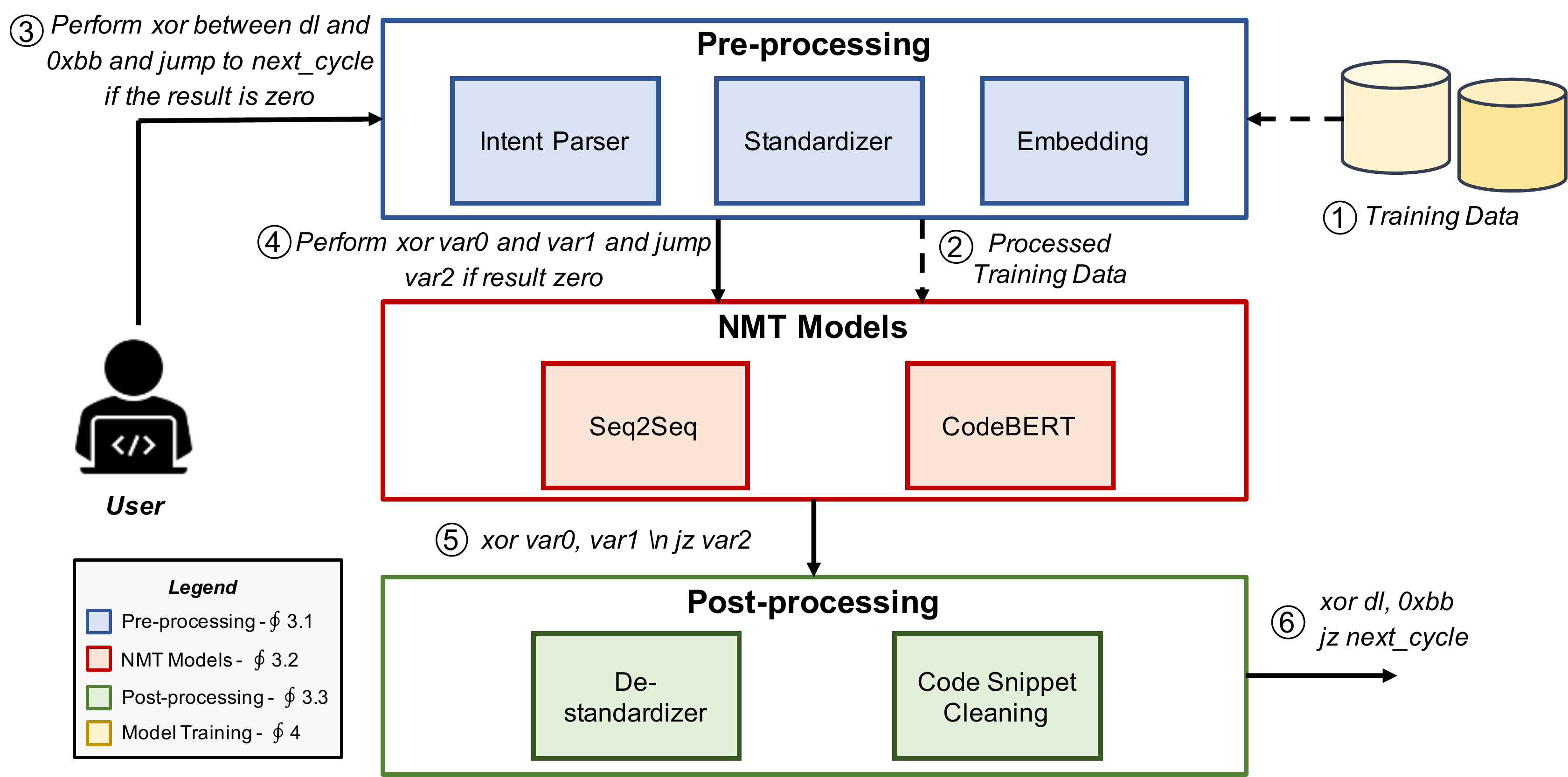}
  \caption{Architecture diagram of \approach{} demonstrating a six-step process: 1) Pre-Processing of the samples in the training data, 2) Training of the neural machine translation system with the processed training data, 3) User stating the operation in the English language (i.e., the \textit{intent}), 4) Pre-processing of the intent, 5) NMT models generating the \textit{code snippet} from the processed intent, and lastly, 6) \approach{} output after post-processing applied to the generated code snippet.}
  \label{fig:approach}
\end{figure*}

The \approach{} approach leverages neural machine translation (NMT) to automatically generate exploits. A neural machine translation system is a neural network that maximizes the conditional probability $p(a|e)$ of translating a source sentence $e = <e_1,...,e_{T_e}>$, with length $|e| = T_e$, into a target sentence $a = <a_1,...,a_{T_a}>$, with length $|a|= T_a$.
Following prior work (e.g.,~\cite{luong2015effective}), we build a neural network that directly models the conditional probability $p(a|e)$ of translating an \emph{intent}, in natural language into a \emph{code snippet} in Python or assembly language.
As a simple example, consider the sequence of English tokens {\fontfamily{qcr} \small\selectfont [`add', `the', `value', `4', `to', `the', `eax', `register']} as $e$, and the sequence of assembly tokens {\fontfamily{qcr} \small \selectfont [`add', `eax', `,', `4']} as $a$, with $|e|=8$ and $|a|=4$. 
The general architecture of an NMT system consists of an encoder, which computes a representation $s$ for each source token, and a decoder, which generates tokens in the target language\footnote{In this section, we use the terms \emph{encoder} and \emph{decoder} to refer to deep learning architectures. In other sections, the terms refer to parts of an exploit.}. 


To support automatic code generation, neural machine translation is usually accompanied by data processing steps \cite{park2020decoding,8713737,oudah2019impact}. These phases strongly depend on the specific source and target languages to translate (in our case, exploit code from the English language). The \approach{} approach applies steps of processing both before the NMT task (\textit{pre-processing}), to train the NMT model and prepare the input data, and after the NMT task (\textit{post-processing}), to improve the quality and the readability of the code in output. 
Figure~\ref{fig:approach} shows the architecture of our approach, along with an example of inputs and outputs at each step, further discussed in the following.

\subsection{Pre-Processing}
\label{subsec:preprocessing}


Pre-processing starts with \textit{stopwords filtering}, i.e., by removing a set of custom compiled words (e.g., \textit{the, each, onto}), in order to include only relevant data for machine translation. 
This phase splits the input sequence of natural language tokens $e_1,...,e_{T_e}$ and code $a_1,...,a_{T_a}$ in a process called \textit{tokenization}. 
The tokenizer converts the input strings into their byte representations, and learns to break down a word into subword tokens (e.g., lower becomes \texttt{[low,er]}. 
We tokenize intents using the \textit{nltk word tokenizer} \cite{loper2002nltk} and snippets using the Python \textit{tokenize} package \cite{tokenize}. We then use regular expressions to identify hexadecimal values (e.g., \texttt{0xbb}), strings that fall between quotation marks, squared brackets, variable name notations (e.g., \texttt{variableName}, \texttt{variable\_name}), follow underscore notation (e.g., \texttt{next\_cycle}), function names, mathematical expressions, and byte arrays (e.g., \texttt{\textbackslash xe3 \textbackslash xa1}). We also use WordNet \cite{miller1995wordnet} to recognize alphabet strings that do not belong to the English language.

Consider the natural language intent shown in Figure~\ref{fig:approach} (step \circled{3}), which contains the phrase \emph{jump to \texttt{next\_cycle}}. 
One task for code generation systems is to prevent non-English tokens (e.g., \texttt{next\_cycle}) from getting transformed during the learning process. This process is known as \textbf{\textit{Standardization}}. Extant code generation systems address this problem with \emph{copying} mechanisms in neural network architectures \cite{gu2016incorporating}, which are inspired by human memorization. 
To perform standardization, \approach{} provides a novel \textbf{\textit{Intent Parser}} tailored for working with exploit software. 
The goal of the Intent Parser is to take input in natural language (i.e. intents) and to provide as output a dictionary of standardizable tokens, such as specific values, label names, and parameters.

All tokens selected by the Intent Parser are passed to the \textbf{\textit{Standardizer}}. The standardization process simply replaces the selected token in both the intent and snippet with \texttt{var\#}, with \texttt{\#} denoting a number from $0$ to $|l|$, and $|l|$ is the number of tokens to standardize.  
In the step~\circled{4} in \figurename~\ref{fig:approach}, the intent parser identifies \texttt{dl}, \texttt{0xbb}, and \texttt{next\_cycle} as standardizable tokens and standardizes them to \texttt{var0}, \texttt{var1}, and \texttt{var2} respectively (based on order of appearance in the intent). 
To prevent the standardization of unimportant tokens, we compile a dictionary of $45$ assembly keywords (e.g., \texttt{register}, \texttt{address}, \texttt{byte}), and $38$ Python keywords (e.g., \texttt{for}, \texttt{class}, \texttt{import}) as non-standardizable tokens. 
After the standardization process, both the original token and its standardized counterpart (\texttt{var\#}) are stored in a dictionary 
to be used during post-processing.

Lastly, we create \textit{\textbf{Embeddings}}, i.e., a mapping of each token (in both the intent and code snippet sequences) into a numerical id representation in order to capture their semantic and syntactic information, where the semantic information correlates with the meaning of the tokens, while the syntactic one refers to their structural roles \cite{li2018word}.

\subsection{NMT Models}
\label{subsec:model}

To perform neural machine translation (step \circled{5}), we consider two standard architectures: Seq2Seq, and CodeBERT.

\textbf{Seq2Seq.} Seq2Seq is a common model used in a variety of neural machine translation tasks. Similar to the encoder-decoder architecture with attention mechanism \cite{bahdanau2014neural}, we use a bi-directional LSTM as the encoder, to transform an embedded intent sequence $e = |e_1,...,e_{T_e}|$ into a vector $c$ of hidden states with equal length. 
Within the bidirectional LSTM encoder, each hidden state $h_t$ corresponds to an embedded token $e_t$. The encoder LSTM is bidirectional, which means it reads the source sequence $e$ ordered from left to right (from $e_1$ to $e_{T_e}$) and from right to left (from  $e_{T_e}$ to $e_1$). To combine both directions, each hidden state for the bidirectional LSTM encoder is computed by concatenating the hidden states of the forward and backward orders at token $t$ as follows:
\begin{equation}
\label{eq:encoding}
\begin{matrix}
\textbf{h}_t = [f(e_t,\textbf{h}_{t-1}) ; f(e_t,\textbf{h}_{t+1})] ;  
\end{matrix}
\end{equation}
where $h_t$ denotes a hidden state at time step $t$, $e_t$ denotes an embedded intent token $t$, and $f$ denotes an LSTM non-linear function. 
Next, the model generates a context vector $c_t$ for each of the embedded inputs.  We use the Bahdanau-style attention mechanism \cite{bahdanau2014neural}, 
which uses soft attention by representing $\textbf{c}_t$ as the weighted sum of the encoder hidden states, as $\mathbf{c}_t = \sum_{i=1}^{{T_e}} \alpha_{t,i}\: \textbf{h}_i$, where the scores $\alpha_{t,i}$ are parametrized by a feed-forward multi-layer perceptron neural network. Since the encoder uses a bidirectional LSTM, each hidden state $\textbf{h}_i$ is aware of the context on both ends.



The decoder generates one target token at a time by decomposing the conditional probability $p(a|e)$ into $\prod_{t=1}^{T_a} p(a_t|a_{<t},s)$, where $a_t$ is an individual output token. 
The target token are generated by combining the LSTM decoder state $s_{t-1}$, the previously-generated token $a_{t-1}$, and the context vector $c_t$ as follows:
\begin{equation}
\label{eq:seq2seq}
\begin{matrix}
\textbf{s}_t = f(a_{t-1},\textbf{s}_{t-1}, \textbf{c}) \\ \\
\: \: \: \: \: \:  p(a_t|a_{<t},e) = g( a_{t-1},\textbf{s}_t,\textbf{c})
\end{matrix}
\end{equation}

\noindent
where $g$ is a non-linear function, and  $a_{<t}$ denotes previous predicted tokens $\{a_1,...,a_{t-1}\}$.





\textbf{CodeBERT.} CodeBERT~\cite{feng_codebert_2020} is a large multi-layer bidirectional Transformer architecture~\cite{vaswani2017attention}. Like Seq2Seq, the Transformer architecture is made up of encoders and decoders. CodeBERT has 12 stacked encoders and 6 stacked decoders. Compared to Seq2Seq, the Transformer architecture introduces mechanisms to address key issues in machine translation: (i) the translation of a word depends on its position within the sentence; (ii) in the target language, the order of the words (e.g., adjectives before a noun) can be different from the order of words in the source language (e.g., adjectives after a noun); (iii) several words in the same sentence can be correlated (e.g., pronouns). These problems are especially important when dealing with long sentences.

The Transformer architecture first refines the input embedding of each token, by combining it with a \emph{positional encoding} vector. The architecture has a different positional encoding vector for each position of the sentence, in order to enrich the input embedding with positional information. Then, the transformed input embeddings sequentially go through the stacked encoder layers, which all apply a \emph{self-attention} process. The self-attention further refines an input embedding, by combining it with the other input embeddings for the sentence in a weighted way, in order to account for correlations among the words (e.g., to get information for a pronoun from the noun it refers to, the input embedding of the noun is given a large weight). The weights are given by:

\begin{equation}
\label{eq:transformers_attention}
Attention (Q,K,V) = softmax(\frac{QK^T}{\sqrt{d_k}})V
\end{equation}

\noindent
where the vectors $Q$ (query), $K$ (key), and $V$ (value) are learned during training. To further improve this mechanism, the Transformer architecture uses \emph{multi-headed attention}, where input embeddings are first multiplied with three learned weight matrices $W_{i}^Q$, $W_{i}^K$, and $W_{i}^V$ (where $i$ is a ``head''), then combined with vectors $Q$, $K$, and $V$. Finally, the results from all heads are concatenated and multiplied with an additional weights matrix $W^O$. Multi-headed attention enables the neural network to correlate different parts of the sequence in different ways (e.g., short-term vs long-term correlations). 

\begin{equation}
\label{eq:transformers_multiattention}
MultiHead (Q,K,V) = Concat(head_1,...,head_h)W^O
\end{equation}



Different from Seq2Seq, CodeBERT also comes with a \emph{pre-trained} neural network model, learned from large amounts of code snippets and their descriptions in the English language, and covering six different programming languages, including Python, Java, Javascript, Go, PHP, and Ruby. The goal of pre-training is to bootstrap the training process, by establishing an initial version of the neural network, to be further trained for the specific task of interest \cite{peters2018deep,liu2019roberta,devlin2018bert,brown2020language}. This approach is called \emph{transfer learning}. In our case, we fine-tune the CodeBERT model to translate English to Python and assembly, using our exploit dataset (see \S{}~\ref{sec:dataset}).



CodeBERT undergoes unsupervised pre-training using two optimization objectives: a \textit{masked language modeling} (MLM) objective~\cite{devlin2018bert} and a \textit{replaced token detection} (RTD) objective ~\cite{clark2020electra}. These two objectives are combined into one loss function $\underset{\theta}{min} \;\; L_{MLM} (\theta) + L_{RTD} (\theta)$.

The MLM objective selects random positions within the intent word sequence $e$ and code snippet sequence $c$, and replaces them with a special mask token $[MASK]$. 
The objective of this pre-training is to predict the original tokens that were replaced with $[MASK]$:

\begin{equation}
\label{eq:transformers_mlm}
L_{MLM} (\theta) = \sum_{i \in m^e \bigcup m^c} - \log p^{D_1} (x_i | e^{masked} , c^{masked})
\end{equation}

\noindent
where $p^{D_1}$ is the model that predicts the masked tokens, $m^e$ and $m^c$ are the randomly selected positions, $x = \{e,c\}$ is the original intent-snippet pair, and $e^{masked}$ and $c^{masked}$ are the masked intent and snippet.

In the RTD objective, both the intent and snippet are corrupted by replacing the original token with incorrect ones, and the trained model is used as a \emph{discriminator} (denoted as $p^{D_2}$) to detect whether a token is original or replaced. To this purpose, CodeBERT uses two \emph{generator} models (similarly to Generative Adversarial Networks), one for intents $p^{G_e}$ and one for snippets $p^{G_c}$. The generator models are trained to generate plausible token fits for the masked tokens $e^{masked}$ and $c^{masked}$. 
The RTD objective is defined as follows:

\begin{equation}
\label{eq:transformers_RTD}
\begin{matrix}
L_{RTD} (\theta) = \mathlarger{\sum}_{i = 1} ^{|e| + |c|} \Big(\sigma(i) \log p^{D_2}  (x_{corrupt}, i) + \big( 1 - \sigma(i)\big)\\ \big(1 - \log p^{D_2} (x^{corrupt} , i ) \big) \Big)\\ \\
\sigma (i) = \Big\{\begin{matrix} 1, \; \;  \mathrm{if} \; x_i^{corrupt} = x_i\\0, \; \; \mathrm{otherwise.}
\end{matrix}
\end{matrix}
\end{equation}


\subsection{Post-Processing}
\label{subsec:postprocessing}
Post-processing is an automatic post-editing process applied during decoding in the translation process.
The \textbf{\textit{Destandardizer}} uses the slot map dictionary generated by the Intent Parser to replace all keys in the standardized intent (i.e., \texttt{var0}, \texttt{var1}  and \texttt{var2}) with the corresponding memorized values (i.e., \texttt{dl}, \texttt{0xbb}, and \texttt{next\_cycle}). 

The generated snippet is then further post-processed using regular expressions (\textbf{\textit{Code Snippet Cleaning)}}. 
This operation includes the removal of (any) extra-spaces in the output, such as between operations and operands, between byte string identifiers and the byte strings, between objects and method calls in Python, etc., and the removal of (any) extra-backslashes in escaped characters (e.g., \texttt{\textbackslash{\textbackslash{n}}}). Also, during the post-processing, the newline characters \texttt{\textbackslash{n}} are replaced with new lines to generate multi-line snippets.
As a final step, snippet tokens are joined to form a complete snippet (step \circled{6}).

\subsection{Implementation Details}
\label{subsec:implementation}

We implement the Seq2Seq model using {\fontfamily{qcr}\selectfont xnmt} \cite{neubig2018xnmt}. We use an Adam optimizer \cite{kingma2014adam} with $\beta_1=0.9$ and $\beta_2=0.999$, while the learning rate $\alpha$ is set to $0.001$. We set all the remaining hyper-parameters in a basic configuration: layer dimension = $512$, layers = $1$, epochs (with early stopping enforced) = $200$, beam size = $5$. 

Our CodeBERT implementation uses an encoder-decoder framework where the encoder is initialized to the pre-trained CodeBERT weights, and the decoder is a transformer decoder. The decoder is composed of $ 6$ stacked layers. The encoder follows the RoBERTa architecture~\cite{liu2019roberta}, with $12$ attention heads,  hidden layer dimension of $768$, $12$ encoder layers, $514$ for the size of position embeddings. We use the Adam optimizer~\cite{kingma2014adam}. The total number of parameters is 125M. The max length of the input is $256$ and the max length of inference is $128$. The learning rate $\alpha = 0.00005$, batch size = $32$, beam size = $10$, and assembly train\_steps = $2800$, and Python train\_steps = $18000$.

%% file: tex/dataset.tex
To automatically generate Python and assembly programs used for security exploits, we curated a large dataset for feeding NMT techniques.
We collected exploits from publicly available databases \cite{exploitdb, shellstorm}, public repositories (e.g., GitHub), and programming guidelines. In particular, we focused on exploits targeting Linux, the most common OS for security-critical network services, running on \textit{IA-32} (i.e., the 32-bit version of the x86 Intel Architecture). The dataset consists of two parts: (i) a Python dataset, which contains Python code used by exploits to encode the shellcode, and (ii) an assembly dataset, which includes shellcode and decoders to revert the encoding. A sample in the dataset consists of a snippet of code from these exploits and their corresponding description in the English language.
The datasets are processed though the operations described in \S{}~\ref{subsec:preprocessing}, and used to train the NMT models, as shown in in \figurename{}~\ref{fig:approach} (steps \circled{1} and \circled{2}).

To deal with the ambiguity of natural language, multiple authors worked together to describe curated snippet intentions in English. The building process of the datasets is similar to established corpora in the NMT field (e.g., the authors of the widespread Django-dataset \cite{oda2015learning} hired one engineer to create the corpus). To mitigate bias, we reused the comments written by developers of the collected programs; when not available, we followed the style of books/tutorials on assembly/Python and shellcode programming.

Table~\ref{tab:dataset_statistics} summarizes the statistics of both datasets, including the size (i.e., the unique pairs of intents-snippets), the unique lines of code snippets, the unique lines of natural language intents, the unique number of tokens (i.e., words), and the average number of tokens per snippet and intent.

\begin{table}[t]
\caption{Datasets statistics}
\label{tab:dataset_statistics}
\centering
\begin{tabular}{
>{\centering\arraybackslash}m{3cm} |
>{\centering\arraybackslash}m{2.25cm}
>{\centering\arraybackslash}m{2.25cm}}
\toprule
\textbf{Statistic} & \textbf{Encoder Dataset} & \textbf{Decoder Dataset}\\ \midrule
\textit{Dataset size} & $15,540$ & $3,715$ \\ 
\textit{Unique Snippets} & $14,034$ &  $2,542$ \\ 
\textit{Unique Intents} & $15,421$ & $3,689$ \\
\textit{Unique tokens (Snippets)} & $9,511$ &  $1,657$ \\ 
\textit{Unique tokens (Intents)} & $10,605$ & $1,924$ \\ 
\textit{Avg. tokens per Snippet} & $11.90$ & $4.75$\\ 
\textit{Avg. tokens per Intent} & $14.90$ & $9.53$\\ \bottomrule
\end{tabular}
\end{table}


\subsection{Python Data}
\label{subsec:encoderdataset}

Our first dataset contains samples to generate Python code for security exploits. In order to make the dataset representative of real exploits, it includes code snippets drawn from exploits from public databases. Differing from general-purpose Python code found in previous datasets \cite{oda2015learning}, the Python code of real exploits entails low-level operations on byte data for obfuscation purposes (i.e., to \emph{encode} shellcodes). Therefore, real exploits make extensive use of Python instructions for converting data between different encoders, for performing low-level arithmetic and logical operations, and for bit-level slicing, which cannot be found in the previous general-purpose Python datasets. 

In total, we built a dataset that consists of $1,114$ original samples of exploit-tailored Python snippets, and their corresponding intent in the English language. These samples include complex and nested instructions, as typical of Python programming. Table~\ref{tab:encoder_instructions} shows examples of such instructions.
In order to perform more realistic training and for a fair evaluation, we left untouched the developers' original code snippets and did not decompose them. We provided English intents to describe nested instructions altogether. 

\begin{table}[t]
\caption{Examples of encoding instructions in Python}
\label{tab:encoder_instructions}
\begin{tabular}
{>{\centering\arraybackslash}m{4cm} | 
>{\centering\arraybackslash}m{4cm}}
\toprule
\textbf{Code Snippet} & \textbf{English Intent} \\ \midrule
\texttt{sb = int(hex(leader)[3:],16)}
& \textit{Convert the value of leader to hexadecimal, then slice it at index 3, convert it to an int16 and set its value to the variable sb}\\ \midrule
\texttt{val2 = int( chunk[i].encode('hex'), 16 ) \^{} xor\_byte}
& \textit{val2 is the result of the bitwise xor between the integer base 16 of the element i of chunk encoded to hex and xor\_byte} \\
\bottomrule
\end{tabular}
\end{table}

In order to bootstrap the training process for the NMT model, we include in our dataset both the original, exploit-oriented snippets and snippets from a previous general-purpose Python dataset. This enables the NMT model to generate code that can mix general-purpose and exploit-oriented instructions. Among the several datasets for Python code generation, we choose the Django dataset \cite{oda2015learning} due to its large size. This corpus contains $14,426$ unique pairs of Python statements from the Django Web application framework and their corresponding description in English.
Therefore, our final dataset contains $15,540$ unique pairs of Python code snippets alongside their intents in natural language.  


\subsection{Assembly Data}
\label{subsec:decoderdataset}

We built our assembly dataset on top of our previous work \cite{liguori-etal-2021-shellcode}, in which we released a dataset for automatically generating assembly from natural language descriptions. 
This dataset consists of \numprint{3,200} assembly instructions, commented in English language, which were collected from shellcodes for \textit{IA-32} and written for the \textit{Netwide Assembler} (NASM) for Linux \cite{duntemann2000assembly}.
In order to make the data more representative of the code that we aim to generate (i.e., complete exploits, inclusive of \emph{decoders} to be delivered in the shellcode), we enriched the dataset with further samples of assembly code, drawn from the exploits that we collected from public databases. Differently from the previous dataset, the new one includes assembly code from \emph{real decoders} used in actual exploits. The final dataset contains $3,715$ unique pairs of assembly code snippets/English intents. 

To better support developers in the automatic generation of the assembly programs, we looked beyond a one-to-one mapping between natural language intents and their corresponding code. 
Therefore, the dataset includes $783$ lines (${\sim}21\%$ of the dataset) of \emph{multi-line intents}, i.e., intents that generate multiple lines of assembly code, separated by the newline character \textit{\textbackslash{n}}. These multi-line snippets contain a number of different assembly instructions that can range between $2$ and $5$. 
For example, the copy of the ASCII string \textit{``/bin//sh"} into a register is a typical operation to spawn a shell, which requires three distinct assembly instructions, as shown by the lines 6-7-8 of Listing~\ref{list:shellcode}: push the hexadecimal values of the words \textit{``/bin"} and \textit{``//sh"} onto the stack register before moving the contents of the stack register into the destination register. 
Further examples of multi-line snippets include conditional jumps, tricks to zero-out the registers without generating null bytes, etc. 
Table~\ref{tab:decoder_instructions} shows two further examples of multi-line snippets with their natural language intents. 

\begin{table}[t]
\caption{Examples of decoding instructions in assembly}
\label{tab:decoder_instructions}
\begin{tabular}
{>{\centering\arraybackslash}m{4cm} | 
>{\centering\arraybackslash}m{4cm}}
\toprule
\textbf{Code Snippet} & \textbf{English Intent} \\ \midrule
\texttt{xor bl, 0xBB} \textbackslash{n} \texttt{jz formatting} \textbackslash{n}
\texttt{mov cl, byte [esi]}
& \textit{Perform the xor between BL register and 0xBB and jump to the label formatting if the result is zero else move the current byte of the shellcode in the CL register.}\\ \midrule
\texttt{xor ecx, ecx} \textbackslash{n} \texttt{mul ecx} &
\textit{Zero out the EAX and ECX registers.
}\\ \bottomrule
\end{tabular}
\end{table}

%% file: tex/experiments.tex
This section presents an extensive evaluation of our approach to generating exploits from natural language descriptions. 
We separate the evaluation of Python code generation (i.e., the encoding part of the exploit) and assembly code generation (i.e., the decoding part of the exploit).
Thus, we use distinct test sets for evaluating Python encoders and assembly language decoders, respectively. 
We trained two distinct models, using respectively the Python and the assembly dataset. The models adopt the same architecture.

Differently from previous work in code generation tasks \cite{DBLP:journals/corr/LingGHKSWB16,yin2017syntactic,barone2017parallel,yin2018tranx,yin2019reranking,Xu2020IncorporatingEK}, we did not randomly sample individual instructions from the dataset when dividing the data between training and test set. Indeed, since the ultimate goal of the programmer is to generate exploits in their entirety, we took all instructions from an exploit as a whole. Our test sets cover $20$ different exploits (i.e., $20$ Python programs, and $20$ assembly programs). 
We exclude \emph{print} statements in the Python source code since they are not actually needed for the exploit. 
The exploits and their encoding/decoding schemes have varying complexity and were developed by different programmers for different purposes.
The average number of lines is $23.4$ (median is $19$) for the Python programs and $26.4$ (median is $24$) for the decoders in assembly language.
The pairs intents-snippets in the test set are unique and are not included in the training/dev sets.
Further information on the test set is described in Appendix\footnote{\url{https://github.com/dessertlab/EVIL}}.

Next, we present the experimental results using automated metrics (\S{}~\ref{subsec:automatic}) and manual metrics (\S{}~\ref{subsec:human}). 
In \S{}~\ref{subsec:exploit_eval}, we evaluate the approach at generating exploits in their entirety. In \S{}~\ref{subsec:times}, we evaluate the computational cost.

\subsection{Automatic Evaluation}
\label{subsec:automatic}

\begin{table*}[ht]
\small
\centering
\caption{Automated evaluation of the translation task. Bolded values are the best performance. IP= Intent Parser}
\label{tab:automatic_evaluation}
\begin{tabular}{
>{\centering\arraybackslash}m{1.5cm}| 
>{\centering\arraybackslash}m{2.5cm}
>{\centering\arraybackslash}m{1.5cm}
>{\centering\arraybackslash}m{1.5cm}
>{\centering\arraybackslash}m{1.5cm}
>{\centering\arraybackslash}m{1.5cm}
>{\centering\arraybackslash}m{1.5cm}}
\toprule
\textbf{Dataset} & \textbf{Model} & \textbf{BLEU-1 (\%)} & \textbf{BLEU-2 (\%)} & \textbf{BLEU-3 (\%)} & \textbf{BLEU-4 (\%)} & \textbf{ACC (\%)} \\ \midrule

\multirow{4}{*}{\textbf{\textit{Python}}}
& \textit{Seq2Seq no IP} & 72.34 & 62.55 & 56.88 & 52.2 & 31.2\\
& \textit{Seq2Seq with IP} & 86.92 & 83.68 & 81.42 & 79.69 & 45.33\\
& \textit{CodeBERT no IP} & 79.23 & 74.12 & 69.84 & 65.76 & 48.00\\ 
& \textit{CodeBERT with IP} & \textbf{89.22} &\textbf{ 86.78} & \textbf{84.94} & \textbf{83.50} & \textbf{56.00}\\
\midrule

\multirow{4}{*}{\textbf{\textit{Assembly}}} 
& \textit{Seq2Seq no IP} & 35.83 & 26.1 & 21.38 & 17.69 & 25.25\\
& \textit{Seq2Seq with IP} & \textbf{88.38} & \textbf{86.37} & \textbf{85.51} & \textbf{85.05} & 40.98\\
& \textit{CodeBERT no IP} & 33.42 & 28.39 & 25.38 & 22.72 & 45.9\\ 
& \textit{CodeBERT with IP} & 88.21 & 85.53 & 84.05 & 82.99 & \textbf{45.9}\\
\bottomrule
\end{tabular}
\end{table*}


We evaluate the translation ability of both models described in \S{}~\ref{subsec:model}, i.e., Seq2Seq with attention mechanisms and CodeBERT. Moreover, since the Intent Parser plays a critical role in our approach, we evaluate the ability of the models with and without the use of the Intent Parser, in order to estimate its contribution.
The configuration without the Intent Parser still adopts the pre/post-processing pipeline (stopwords filtering, tokenization, embeddings, etc.) but avoids standardization.

Automatic evaluation metrics are commonly used in the field of machine translation. They are reproducible, easy to be tuned, and time-saving.
The \emph{BiLingual Evaluation Understudy} (BLEU) \cite{papineni2002bleu} score is one of the most popular automatic metric \cite{oda2015learning,DBLP:journals/corr/LingGHKSWB16,gemmell2020relevance,tran2019does}. This metric is based on the concept of \textit{n-gram}, i.e., the adjacent sequence of $n$ \textit{items} (e.g., syllables, letters, words, etc.) from a given example of text or speech. In particular, this metric measures the degree of n-gram overlapping between the strings of words produced by the model and the human translation references at the corpus level. BLEU measures translation quality by the accuracy of translating n-grams to n-grams, for n-gram of size $1$ to $4$ \cite{han2016machine}.
The \emph{Exact match accuracy} (ACC) is another automatic metric often used for evaluating neural machine translation  \cite{DBLP:journals/corr/LingGHKSWB16,yin2017syntactic,yin2018tranx,yin2019reranking}. It measures the fraction of the exact match between the output predicted by the model and the reference.

Table~\ref{tab:automatic_evaluation} shows the best results for each model in terms of BLEU scores and accuracy. 
For both the Python and assembly datasets, the results point out that the Intent Parser notably increases the performance of the translation task in Python and assembly. Moreover, CodeBERT outperforms the performance of the Seq2Seq with respect to all metrics for the Python dataset. In the case of the assembly dataset, the performance of the two models is comparable.

\subsection{Human Evaluation}
\label{subsec:human}

We further investigate the performance of the translation task by assessing the deeper linguistic features \cite{DBLP:journals/corr/abs-2105-03311}, i.e., the syntax and the semantic of the code snippets predicted by the models. These features allow us to properly give credit to semantically-correct code that fails to match the reference one (e.g., \texttt{jz label} and \texttt{je label} are semantically identical code snippets, even if they use different instructions), or to provide information whether the generated code would compile or not.
Accordingly, we define two new metrics: a generated output snippet is considered \textbf{\textit{syntactically correct}} if it is correctly structured in the grammar of the target language (i.e., Python or assembly) and compiles correctly.
The output is considered \textbf{\textit{semantically correct}} if the snippet is an appropriate translation in the target language given the intent description. The semantic correctness implies syntax correctness, while a snippet can be syntactically correct but semantically incorrect. Of course, the syntactic incorrectness also implies the semantic one. We evaluated these metrics through manual inspection.

As a simple example, consider the intent \textit{res2 is the result of the bitwise and operation between res2 and val1}. If the model generates  \texttt{res2 = res2 \& val1}, then the snippet is considered semantically and syntactically correct. 
If the model generates the Python instruction \texttt{res2 = val2 | val1}, then the snippet is considered syntactically correct for the Python syntax, but semantically incorrect both because the bitwise operation is not the intended one (i.e., the \texttt{or} instead of the \texttt{and}), and the operands are not the same specified in the intents (i.e., \texttt{val2} instead of \texttt{res2}). If the model generates the instruction \texttt{res2 = res2 \_ val1}, then the snippet is considered also syntactically incorrect, since the symbol \texttt{\_} is not a valid binary operator in Python. 
Notice that this type of evaluation is rigorous. Even if a single token of the generated snippet (e.g., an operation, a variable/operand, a parenthesis, etc.) is not syntactically (semantically) correct, then the whole snippet is considered syntactically (semantically) incorrect.
In the case of the multi-line snippets of the assembly dataset, we compute the syntactic (semantic) correctness as the ratio number of syntactically (semantically) correct single snippets over the total number of snippets composing the multi-lines statement (c.f. \S{}~\ref{subsec:failure_cases}).  


\begin{table}[t]
\small
\centering
\caption{Human evaluation of the translation task. Bolded values are the best performance  ($*=$ p$<$0.05). IP= Intent Parser}
\label{tab:human_evaluation}
\begin{tabular}{
>{\centering\arraybackslash}m{1.5cm}| 
>{\centering\arraybackslash}m{1.5cm}
>{\centering\arraybackslash}m{2cm}
>{\centering\arraybackslash}m{2cm}
}
\toprule
\textbf{Dataset} & \textbf{Model} & \textbf{Syntactic Correctness (\%)} & \textbf{Semantic Correctness (\%)} \\ \midrule

\multirow{2}{*}{\textbf{\textit{Python}}} 
& \textit{Seq2Seq with IP} & 88.53 & 50.13\\
& \textit{CodeBERT with IP} & \textbf{93.60*} & \textbf{67.73*}\\
\midrule
\multirow{2}{*}{\textbf{\textit{Assembly}}} 
& \textit{Seq2Seq with IP} & \textbf{90.16} & 56.36\\
& \textit{CodeBERT with IP} & 87.05 & \textbf{61.90*}\\
\bottomrule
\end{tabular}
\end{table}



We evaluated both types of correctness achieved by Seq2Seq and CodeBERT on the entire test-sets (all 20 encoders/decoders). For the evaluation, we were supported by the Python 2 and the NASM compiler, respectively for Python and assembly code. 
Since the previous analysis showed that the Intent Parser notably increases the performance of translation, we evaluated the syntactic and semantic correctness only when using the Intent Parser. 
Table~\ref{tab:human_evaluation} shows the percentage of the syntactically and semantically correct snippets generated by the models. The table shows that, for the Python programs, CodeBERT with IP provides a higher percentage of both syntactically and semantically correct snippets. 
We conducted a \textit{paired-sample T-test} to compare the syntactic correctness and the semantic correctness values of the code snippet pairs predicted by the two NMT models (given the same intents). We found that the differences between CodeBERT and the Seq2Seq are statistically significant for both metrics with $p < 0.05$.
For the assembly programs, Seq2Seq provides a higher percentage of syntactically correct snippets, but these differences are not statistically significant. Again, CodeBERT outperforms Seq2Seq in the semantic correctness ($p < 0.05$).

These findings highlight that CodeBERT predicts the highest percentage of code snippets that are semantically equivalent to the English intent. It is interesting that, differently from the Seq2Seq, this model provides better performance when applied to the Python dataset. We attribute these differences to the pre-training of the model, which benefits from knowledge from six high-level programming languages, including Python. 
Finally, we investigated the difference between the percentage of syntactic correctness and the semantic one. We found that most of the semantically incorrect predictions are due to wrong labels or variable names, or the omission of parenthesis around expressions. These errors are easy to identify and correct by a programmer, as they would require one single edit to make the predicted snippet semantically consistent with the intent. These errors are further analyzed in \S{}~\ref{sec:qualitative}.

According to these results, the NMT approach can correctly translate an individual intent with high likelihood, and the incorrect translations are still close to being correct. These results support the use of the NMT approach as a way for developers to look up code snippets that they could not recall or that are not confident yet to develop themselves, getting a close-to-correct snippet that they can use with little effort.

\subsection{Whole-exploit evaluation}
\label{subsec:exploit_eval}

The ultimate goal of the programmer is to generate entire software exploits. While evaluation using the manual metrics indicates that the model can correctly generate individual snippets with high likelihood, we do not know yet whether the model can generate a correct exploit as a whole. 
Therefore, we evaluated the ability of the approach to generate semantically and syntactically correct code for entire software exploits (i.e., all of the lines of code in the programs), using two new metrics.

Let $n_t^i$ be the the number of total lines of the $i$-th program in the test set ($i \in [1,20]$). Let also consider $n_{syn}^i$ as the number of automatically-generated snippets for the $i$-th program that are syntactically correct, and $n_{sem}^i$ as the number of automatically-generated snippets that are semantically correct. For every program of the test set, we define the \textit{\textbf{syntactic correctness}} of the program $i$ as the ratio $n_{syn}^i / n_t^i$, and the \textit{\textbf{semantic correctness}} of the program as the ratio $n_{sem}^i / n_t^i$. Both metrics range between $0$ and $1$.

\begin{figure}[t]
    \centering
    \includegraphics[width=0.75\columnwidth]{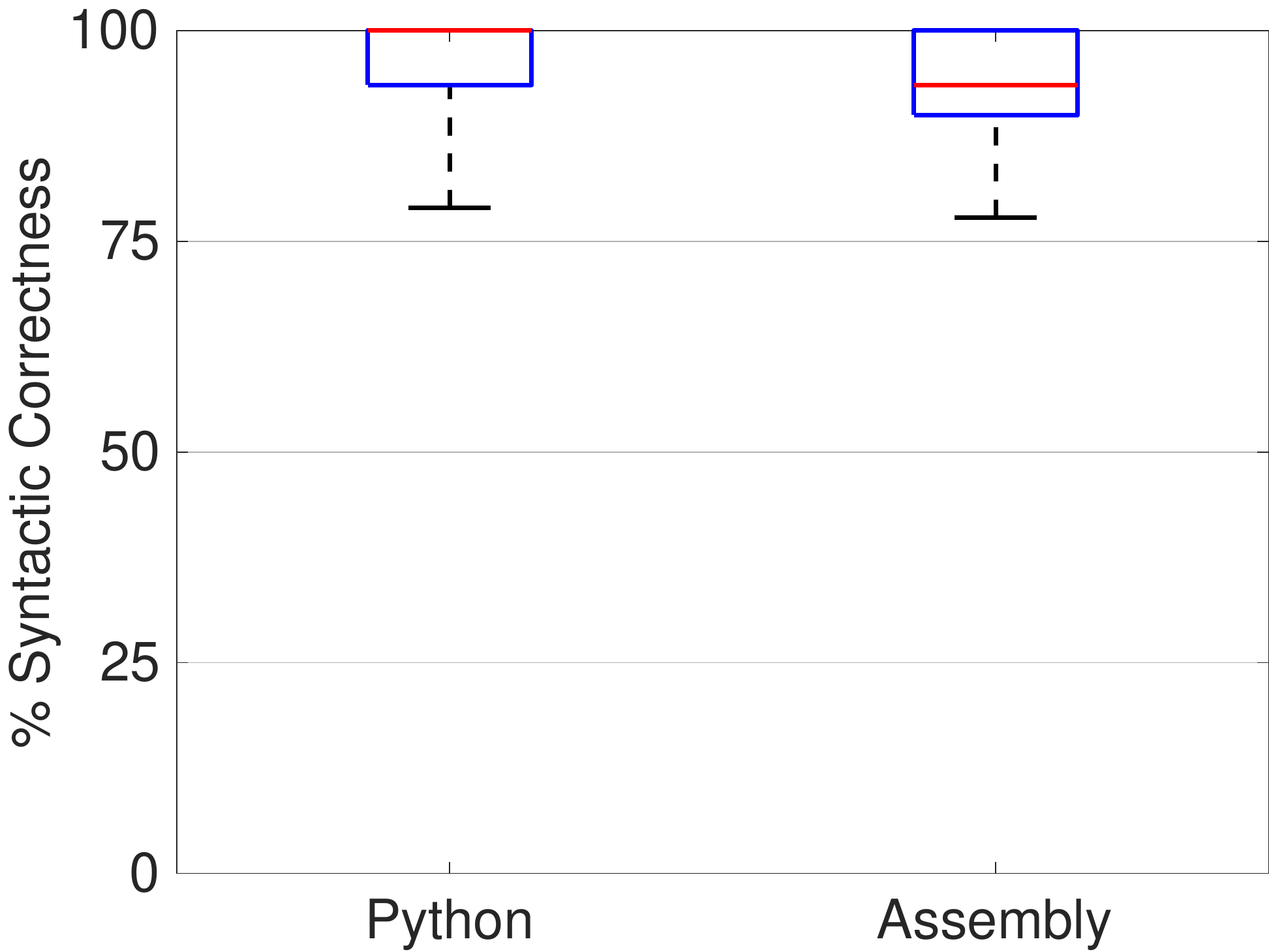}
    \caption{Distribution of the syntactic correctness of the programs.}
    \label{fig:syntax_correc}
\end{figure}

\begin{figure}[t]
    \centering
    \includegraphics[width=0.75\columnwidth]{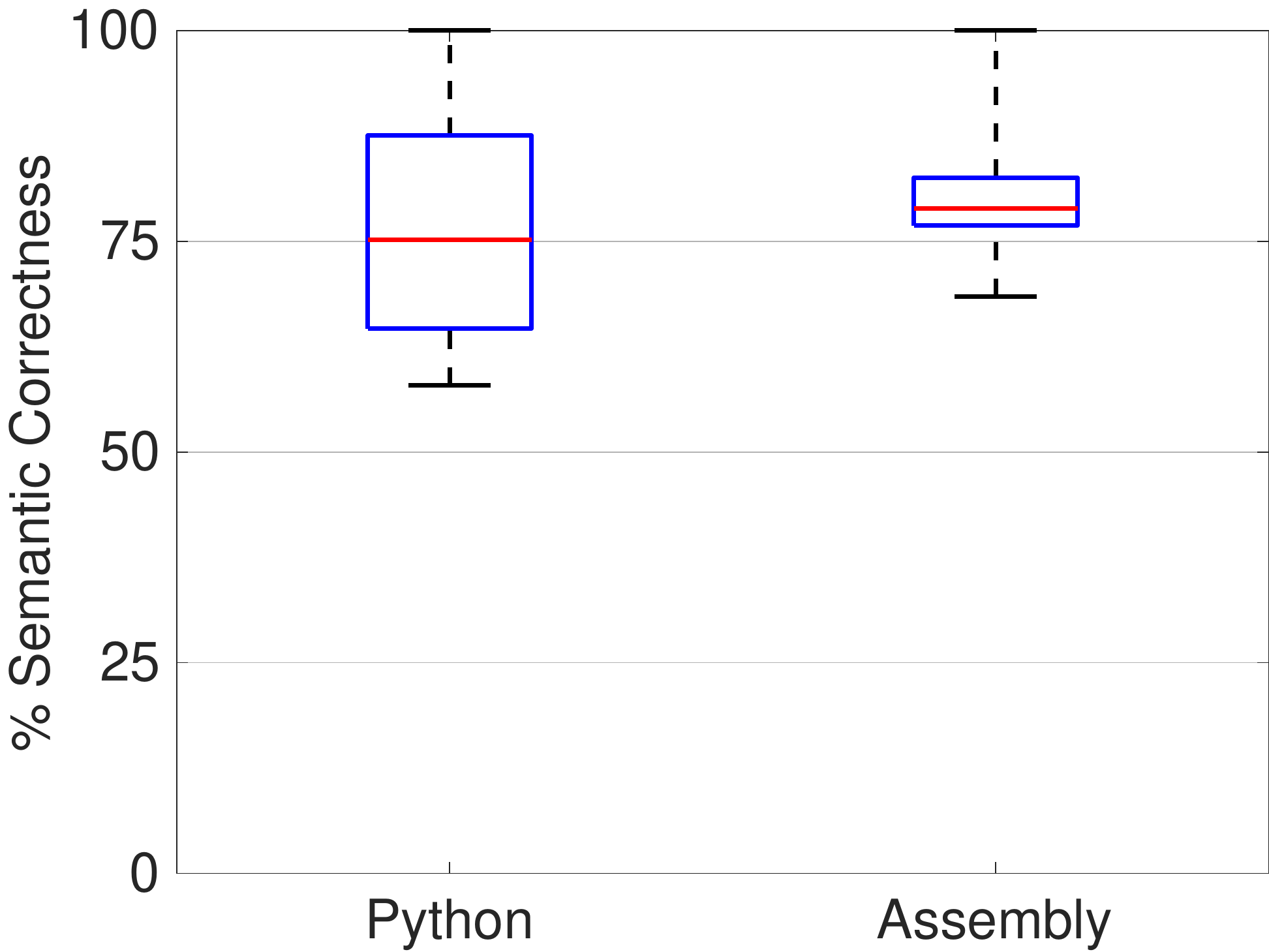}
    \caption{Distribution of the semantic correctness of the programs.}
    \label{fig:semantic_correc}
\end{figure}

Therefore, $\forall i \in [1,20]$, we computed the values $n_{syn}^i$ and $n_{sem}^i$ for Python and assembly programs. We performed this analysis with CodeBERT with IP since it showed the best performance for both datasets in the human evaluation (\S{}~\ref{subsec:human}).
We found that the average syntactic correctness over all the programs of the test set is ${\sim}96\%$ for the Python programs and ${\sim}93\%$ for the assembly programs. Similarly, we estimated the average semantic correctness, which is equal to $\sim76\%$ and ${\sim}81\%$ for Python and assembly programs, respectively. 
The box-plots in \figurename{}~\ref{fig:syntax_correc} and \figurename{}~\ref{fig:semantic_correc} summarize the results.

Despite our conservative evaluation, the approach is able to generate $12$ Python programs and $6$ assembly programs fully composed by syntactically correct snippets (i.e., the $n_{syn}^i / n_t^i = 100\%$).
However, this does not imply that the programs are also executable. For example, the assembly instruction \texttt{jz shellcode} is syntactically correct, but if the label \texttt{shellcode} is not defined, then the program would not compile. Therefore, we evaluated how many programs can be compiled, finding that the $50\%$ of syntactically correct programs are also compilable and executable.
Furthermore, one Python program and one assembly program are fully composed of semantically correct snippets (i.e., $n_{sem}^i / n_t^i = 100\%$) and, therefore, implement the original encoding and decoding schemes.
We interpret these results as a promising \textit{first} research step towards automatically coding software exploits from natural language intents. 

\begin{table*}[ht]
\caption{Illustrative examples of correct output.}
\label{tab:success_cases}
\footnotesize
\centering
\begin{tabular}{
>{\centering\arraybackslash}m{1cm} |
 >{\centering\arraybackslash}m{6.4cm}|
 >{\centering\arraybackslash}m{4.5cm}|
 >{\centering\arraybackslash}m{4.5cm}}
\toprule
\textbf{Dataset} & \textbf{Natural Language Intent} & \textbf{Ground Truth} & \textbf{Model Output}\\ \midrule
\multirow{2}{*}[-0.25cm]{\textbf{\textit{Python}}} & \textit{Take the absolute value of subfs then convert subfs to an integer, then cast to a hexadecimal, slice the variable rev\_suplx between the indicies 0 and 2 then cast rev\_suplx to the type int16, store the value of the summation in the variable xxx} & \texttt{xxx = hex(int(abs(subfs)) + int(rev\_suplx[0:2],16))} & \texttt{xxx = hex(int(abs(subfs)) + int(rev\_suplx[0:2],16))} \\ \cmidrule(l){2-4} 
& \textit{Append 0xAA to the string encode} & \texttt{encode += `0xAA'} & \texttt{encode += `0x\%02x'\%170} \\ \midrule
\multirow{2}{*}[-0.25cm]{\textbf{\textit{Assembly}}} & \textit{Jump short to decode if al is not equal to cl else jump to shellcode.} & \texttt{cmp al, cl \textbackslash n jnz short decode \textbackslash n jmp shellcode} & \texttt{cmp al, cl \textbackslash n jne short decode \textbackslash n jmp shellcode} \\ \cmidrule(l){2-4} 
& \textit{Perform the xor between the current byte of the shellcode and the dl register.} & \texttt{xor byte [esi], dl} & \texttt{xor byte [esi], dl} \\ \bottomrule
\end{tabular}
\end{table*}

\begin{table*}[ht]
\caption{Illustrative examples of incorrect output. The prediction errors are \textcolor{red}{\textbf{red/bold}}. \textcolor{red}{\textbf{\cancel{Slashed}}} text refers to omitted predictions.}
\label{tab:failure_cases}
\footnotesize
\centering
\begin{tabular}{
>{\centering\arraybackslash}m{1cm} |
 >{\centering\arraybackslash}m{6.4cm}|
 >{\centering\arraybackslash}m{4.5cm}|
 >{\centering\arraybackslash}m{4.5cm}}
\toprule

\textbf{Dataset} & \textbf{Natural Language Intent} & \textbf{Ground Truth} & \textbf{Model Output}\\ \midrule
\multirow{2}{*}[-0.5cm]{\textbf{\textit{Python}}} & \textit{Return the result of bitwise left shift of n by rotations bitwise and with mask\_value bitwise or n  bitwise right shift by the subtraction of rotations from width} & \texttt{return ((n<<rotations) \& mask\_value) | (n >> (width $-$ rotations))} & \texttt{return (\textcolor{red}{\textbf{\cancel{(}}}n<<rotations\textcolor{red}{\textbf{\cancel{)}}} \& \textcolor{red}{\textbf{bitwise\cancel{)}}}  | \textcolor{red}{\textbf{\cancel{(}}}n >> \textcolor{red}{\textbf{\cancel{(}}}width $-$ rotations\textcolor{red}{\textbf{\cancel{)}}})} \\ \cmidrule(l){2-4} 
& \textit{Convert the value of leader to hexadecimal, then slice it at index 3, convert it to an int16 and set its value to the variable sb} & \texttt{sb = int(hex(leader)[3:],16)} & \texttt{sb = int(hex(leader)[3:],\textcolor{red}{\textbf{32}})} \\ \midrule
\multirow{2}{*}[-0.5cm]{\textbf{\textit{Assembly}}} & \textit{Zero out the ecx register and move 25 in the lower 8 bits of the register.} & \texttt{xor ecx, ecx \textbackslash n mov cl, 25} & \texttt{xor ecx, ecx \textbackslash n mov \textcolor{red}{\textbf{al}}, 25} \\ \cmidrule(l){2-4} 
& \textit{In the decode function jump to lowbound label if the current byte of the shellcode is lower than 0x7 else subtract 0x7 from the byte of the shellcode and jump to common\_commands} & \texttt{decode: \textbackslash n cmp byte [esi], 0x7 \textbackslash n jl lowbound \textbackslash n sub byte [esi], 0x7 \textbackslash n jmp common\_commands} & \texttt{decode: \textbackslash n cmp byte [esi], 0x7 \textbackslash n jl lowbound \textcolor{red}{\cancel{\textbf{\textbackslash n}}} \textcolor{red}{\cancel{\textbf{sub byte [esi], 0x7}}} \textbackslash n jmp common\_commands} \\ \bottomrule
\end{tabular}
\end{table*}

\subsection{Computational Time}
\label{subsec:times}

We performed our experiments on a Linux OS running on a virtual machine. Seq2Seq utilized $8$ CPU cores and $8$ GB RAM. CodeBERT utilized 8 CPU cores, 16 GB RAM, and 2 GTX1080Ti GPUs.

The computational time needed to generate the output depends on the settings of the hyper-parameters and the size of the dataset. 
On average, the training time of the Seq2Seq model is ${\sim}500$ minutes for the Python dataset and ${\sim}60$ minutes for the assembly dataset, while CodeBERT requires for the training in average ${\sim}220$ minutes for Python and ${\sim}25$ minutes for the assembly dataset. 
In total, we performed $50$ different experiments for a total computational time of ${\sim}150$ hours.
Once the models are trained, the time to predict the output is below 1 second and can be considered negligible. 

%% file: tex/qualitative.tex
In this section, we present a qualitative analysis using cherry-picked examples from our test sets to highlight both successful prediction cases and failed ones.

\subsection{Successful Examples}
\label{subsec:sucessful_cases}

Table~\ref{tab:success_cases} shows four cases of success predictions for both datasets.
The first row demonstrates our approach's ability to generate a difficult Python snippet from a nested natural language intent without any errors. Indeed, the predicted output contains the correct variable name \texttt{subfs}, \texttt{rev\_suplx}, and \texttt{xxx} with the appropriate functions \texttt{hex}, \texttt{int}, \texttt{abs} indexed in the right order with the correct parameters.
The second row shows an example of implicit model knowledge.  The model is able to append \texttt{0xAA} by converting the value 170, the hexadecimal value of \texttt{0xAA}, to hexadecimal and appending it to the string variable encode.

The last two rows of the table show the ability of the model in generating challenging assembly instructions. 
The third row highlights both the ability of the Intent Parser in identifying all the registers' name (i.e., \texttt{AL} and \texttt{CL}), and the labels (i.e., \texttt{decode} and \texttt{shellcode}), and the ability of the model to perform the right translation of the order in the \textit{if-then-else} statements described in the English intent. 
The last row is an interesting example of implicit model knowledge. Indeed, even if it is not stated in the intent, the model is able to properly predict \texttt{EDI} as the destination register since it is typically used to store the encoded shellcodes.

\subsection{Failure Examples}
\label{subsec:failure_cases}


Table~\ref{tab:failure_cases} shows four relevant examples of failure cases. We observe nested instructions within the intent in the Python examples. In the first row, the intent implies the order of operations signified by parenthesis in the ground truth snippet. While the model correctly generates all the operations in the correct order, it does not enforce the order of operations. The model fails to derive the implied order of operations from the intent. It also fails to generate the correct variable \texttt{mask\_value}, but instead generates the intent-repeated token \texttt{bitwise} as a variable, likely due to \texttt{bitwise} being mentioned frequently in the intent in between operations. In the second row, we observe the model also generates all the operands and variables correctly, however it fails to generate the correct parameter 16 to the integer conversion operation.

The example in the third row illustrates an example of failure due to a lack of implicit knowledge. The natural language intent does not mention to the \texttt{CL} register but implicitly refers to it (\textit{lower 8 bits of the \texttt{ECX} register}). Therefore, the output results in a syntactically correct but semantically incorrect prediction.
The last row shows a long intent that describes an entire function. The model is able to properly predict four of the five assembly instructions composing the decode function. However, it misses the \texttt{sub} operation after the conditional jump \texttt{jl}. In this case, the semantic correctness is equal to $0.8$ ($4$ out of $5$ semantically correct snippets).

%% file: tex/related.tex
The task of exploit generation via automatic techniques has been addressed in several ways. 
\textit{ShellSwap} \cite{bao2017your} is a system that generates new exploits based on existing ones, by modifying the original shellcode with arbitrary replacement shellcode. Hu \emph{et al.} \cite{hu2015automatic} developed a novel approach to construct data-oriented exploits through data flow stitching, by composing the benign data flows in an application via a memory error. They built a prototype attack generation tool that operates directly on Windows and Linux x86 binaries. 
Avgerinos \textit{et al.} \cite{avgerinos2011aeg} developed an end-to-end system for automatic exploit generation (AEG) on real programs by exploring execution paths.  Given the potentially buggy program in source form, their proposal automatically looks for bugs, determines whether the bug is exploitable, and produces a working control-flow hijack exploit string. 
\textit{SemFuzz} \cite{you2017semfuzz} extracts necessary information from non-code text related to a vulnerability, using natural language processing and a semantics-based fuzzing process, in order to discover and trigger deep bugs. 
Chen \textit{et al.} \cite{chen2011automatic} presented techniques to find out the \textit{gadgets}, i.e., the basic building block in Jump Oriented Programming (JOP), and showed these gadgets are Turing complete. They implemented an automatic tool able to generate JOP shellcodes.
Ding \textit{et al.} \cite{6999408} proposed a reverse derivation of a transformation method driven by state machines indicating the status of data flows, in order to transform the original shellcode into printable Return Oriented Programming (ROP) payload. 
\textit{Chainsaw} \cite{alhuzali2016chainsaw} is a tool for analyzing web applications and generating injection exploits. The tool performs static analysis and defines a model of the application behavior to generate injection exploits, by leveraging application workflow structures and database schemes.
Brumley \textit{et al.} \cite{4531150} proposed an approach for Automatic Patch-based Exploit Generation (APEG). Starting from a program and its patched version, the approach identifies the security checks added by the patch and automatically generates inputs to fail the checks.
Huang \textit{et al.} \cite{6717039} introduced a method to automatically generate exploits based on software crash analysis. This method analyzes software crashes using a symbolic failure model, to generate exploits from crash inputs and existing exploits for several types of applications.
Xu \textit{et al.} \cite{8432013} developed a tool to find buffer overflow vulnerabilities in binary programs and automatically generate exploits using a constraint solver. Vulnerability detection is achieved through symbolic execution and the exploit generated by this tool can bypass different types of protection.

Our work is radically different from these previous ones. First, our approach uses natural language statements to generate exploits. Second, we adopt neither a static nor dynamic program analysis approach (e.g., fuzzing, program synthesis, etc.), but a statistical, data-driven approach. Therefore, our work can be considered complementary to these previous solutions, with different use cases.

%% file: tex/ethics.tex
\emph{Offensive security} is a sub-field of security research that tests security measures from an adversary or competitor’s perspective, employing ethical hackers to probe a system for vulnerabilities~\cite{bratus2013offensive,oakley2019state}. 
Our work aims to automate exploit generation, in order to explore critical vulnerabilities before they are exploited by attackers~\cite{avgerinos2011aeg}. 
Indeed, our work simplifies the process of coding the exploits to surface security weaknesses within the software and can provide valuable information about the technical skills, degree of experience, and intent of the attackers. With this information, it is possible to implement measures to detect and prevent attacks \cite{arce2004shellcode}.

%% file: tex/conclusion.tex
We presented \approach{}, an approach for automatic exploit generation for security assessment purposes, using natural language processing techniques based on neural networks.
Our approach represents the first step towards the ambitious goal of automatically generating software exploits from natural language. 
We develop and released two datasets of real exploits in Python and assembly language to enable neural network training and experimental evaluation.
We evaluated the feasibility of our approach, using both automated and manual metrics. 
Our experiments have shown the ability of the approach in generating software exploits from natural language descriptions with high syntactic and semantic correctness. 

The results have also revealed that, in most cases, the generated programs do not execute correctly due to wrong labels or variable names. Programmers can easily correct these problems, but our goal is full automation. Therefore, future work includes the improvement of the post-processing phase that looks at the program context to increase the accuracy of the program generation task. 
Future work also includes the development of a single engine that generates the encoding and decoding schemes at the same time, without performing two separate translation tasks.

%% file: tex/appendix_table.tex
\begin{table*}[ht]
\caption{The 20 exploits used as evaluation in test sets. 
\textbf{$n_t$}: number of total lines of the program.
\textbf{$n_{syn}$}: number of syntactically correct lines generated by the approach. 
\textbf{$n_{sem}$}: number of semantically correct lines generated by the approach.}
\label{tab:test_set}
\centering
{\scriptsize
\begin{tabular}
{>{\centering\arraybackslash}m{0.5cm}| 
>{\centering\arraybackslash}m{9cm} |
>{\centering\arraybackslash}m{0.7cm}
>{\centering\arraybackslash}m{0.7cm}
>{\centering\arraybackslash}m{0.7cm} |
>{\centering\arraybackslash}m{0.7cm}
>{\centering\arraybackslash}m{0.7cm}
>{\centering\arraybackslash}m{0.7cm}
}
\toprule
\textbf{id} & \textbf{URL} & \multicolumn{3}{c|}{\textbf{Encoder}} & \multicolumn{3}{c}{\textbf{Decoder}} \\
& & \textbf{$n_{t}$} & \textbf{$n_{syn}$} & \textbf{$n_{sem}$} & \textbf{$n_{t}$} & \textbf{$n_{syn}$} & \textbf{$n_{sem}$}\\ \midrule
1 & \url{https://www.exploit-db.com/shellcodes/47564} & 11 & 11 & 10 & 17 & 17 & 14\\
2 & \url{https://www.exploit-db.com/shellcodes/47461} & 19 & 19 & 17 & 32 & 31 & 25\\
3 & \url{https://www.exploit-db.com/shellcodes/46994} & 21 & 21 & 18 & 27 & 23 & 23\\
4 & \url{https://www.exploit-db.com/shellcodes/46519} & 11 & 11 & 9 & 22 & 20 & 17\\
5 & \url{https://www.exploit-db.com/shellcodes/46499} & 9 & 9 & 8 & 16 & 16 & 14\\
6 & \url{https://www.exploit-db.com/shellcodes/46493} & 9 & 9 & 8 & 16 & 16 & 13\\
7 & \url{https://www.exploit-db.com/shellcodes/45529} & 19 & 15 & 11 & 32 & 32 & 25\\
8 & \url{https://www.exploit-db.com/shellcodes/43890} & 20 & 17 & 16 & 23 & 23 & 22\\
9 & \url{https://www.exploit-db.com/shellcodes/37762} & 26 & 25 & 17 & 24 & 22 & 19\\
10 & \url{https://www.exploit-db.com/shellcodes/37495} & 15 & 14 & 9 & 19 & 17 & 13\\
11 & \url{https://www.exploit-db.com/shellcodes/43758} & 14 & 14 & 9 & 29 & 27 & 24\\
12 & \url{https://www.exploit-db.com/shellcodes/43751} & 8 & 8 & 8 & 46 & 41 & 35\\
13 & \url{https://rastating.github.io/creating-a-custom-shellcode-encoder/} & 64 & 61 & 45 & 27 & 23 & 20\\
14 & \url{https://voidsec.com/slae-assignment-4-custom-shellcode-encoder/} & 18 & 18 & 12 & 18 & 14 & 14\\
15 & \url{https://snowscan.io/custom-encoder/#} & 48 & 45 & 33 & 42 & 38 & 33\\
16 & \url{https://github.com/Potato-Industries/custom-shellcode-encoder-decoder} & 38 & 38 & 32 & 19 & 19 & 19\\
17 & \url{https://medium.com/@d338s1/shellcode-xor-encoder-decoder-d8360e41536f} & 29 & 25 & 25 & 33 & 31 & 25\\
18 & \url{https://www.abatchy.com/2017/05/rot-n-shellcode-encoder-linux-x86} & 10 & 10 & 6 & 17 & 16 & 13\\
19 & \url{https://xoban.info/blog/2018/12/08/shellcode-encoder-decoder/} & 39 & 36 & 23 & 24 & 22 & 19\\
20 & \url{http://shell-storm.org/shellcode/files/shellcode-902.php} & 40 & 40 & 26 & 45 & 44 & 36\\
 \bottomrule
\end{tabular}
}
\end{table*}

Table~\ref{tab:test_set} presents detailed information on the $20$ encoders and decoders in our test sets.
It includes the source URL, the number of total lines ($n_t$) of the programs, and the number of syntactically correct ($n_{syn}$) and semantically correct ($n_{sem}$) lines generated by our approach, for both the encoders in Python and decoders in Assembly. 
In total, the test set for the Python programs contains $375$ unique pairs of Python code snippets (not including \texttt{print}s) along with their natural description. The test set for assembly contains $305$ unique pairs of code snippets ($95$ are multi-line snippets) and natural language intents.